\documentclass[useAMS,usenatbib]{mn2e}
\usepackage{graphicx}

\title[]{The Red Giant Branch in Near\--Infrared Colour\--Magnitude Diagrams. 
	I: The calibration of photometric indices 
\thanks{Based on data taken at the ESO-MPI 2.2m Telescope
	equipped with the near-IR camera IRAC2-ESO, La Silla (Chile),
	within the observing program 59.E-0340.}}

\author[Valenti, Ferraro \& Origlia]{E. Valenti$^{1,2}$, F.~R. Ferraro$^1$, 
        L. Origlia$^{2}$    \\
 $^1$ Dipartimento Astronomia, Universit\`a di Bologna,  
      Via Ranzani 1, I-40127 Bologna, Italy ,\\
      e-mail elena.valenti2@studio.unibo.it, ferraro@bo.astro.it\\
 $^2$ INAF-Osservatorio Astronomico di Bologna, Via Ranzani 1, I-40127 Bologna,
      Italy, \\
      e-mail origlia@bo.astro.it \\
       }
\date{\today}

\begin{document}
\pagerange{\pageref{firstpage}--\pageref{lastpage}} \pubyear{2003}
\maketitle
\label{firstpage}

\begin{abstract}
We present new high\--quality near\--infrared photometry of  
10 Galactic Globular Clusters spanning a wide metallicity range
($-2.12{\leq}{\rm [Fe/H]}{\leq}-0.49$):
 five clusters belong to the Halo (namely, NGC~288,
NGC~362, NGC~6752, M~15 and M~30) and five (namely, NGC~6342,
NGC~6380, NGC~6440, NGC~6441 and NGC~6624) to the Bulge.  
By combining J, H and K observations
with optical data, we constructed Colour\--Magnitude Diagrams in various planes
({\rm (K,J\--K)}, {\rm (K,V\--K)}, {\rm (H,J\--H)}, and 
{\rm (H,V\--H)}).
A set of photometric indices (colours, magnitudes and slopes) 
describing the location and the morphology of the
Red Giant Branch (RGB) have been measured.
We have combined this new data set with those already published by
\citet{F00} and \citet{io}, and here we present an updated calibration
of the various RGB indices in the 2MASS 
photometric system, in terms of the cluster metallicity.
\end{abstract}


\section{Introduction}
The study of 
stellar evolutionary sequences finds several applications in astrophysics:  
inferring the age and metallicity of stellar systems, 
 synthesizing integrated spectra of galaxies, calibrating standard candles 
for distance
determinations. There is a small number of physical
observables that models can predict and that can be compared with observed
quantities. Within this framework, Colour\--Magnitude Diagram (CMD) 
and Luminosity Function (LF) are the most powerful tools to test theoretical
models, being related to the stellar effective temperature, luminosity 
and the duration of a specific evolutionary phase \citep{RF88}. 
In this contest, our group started a long\--term project
devoted to analizing and testing each individual evolutionary sequence in
the CMD of Galactic Globular Clusters (GGCs) \citep[see e.g.][ hereafter
F99 and F00, respectively]{F99,F00}. 
In particular, CMDs and LFs in the near\--Infrared (IR) are useful
in order to perform a detailed study of the Red Giant Branch (RGB). In fact,
in studying cool stellar populations (i.e. RGB stars), the near\--IR spectral
domain offers severals advantages, being the most sensitive to low temperature.
Moreover, 
the background contamination by Main Sequence (MS) stars is much less severe, 
thus allowing to properly characterize the RGB even in the innermost core 
region of stellar clusters affected by crowding.
In addiction, with respect to the visual range, in the IR range the reddening 
is much lower and in some cases, when the extinction is very large, as in the 
Bulge, it represents the only possibility to observe the 
stellar population along the entire RGB.
This is well know since two decades, and several authors have used IR
photometry to derive the main RGB properties 
(see e.g F00 and references therein).\\
By combining near\--IR and optical photometry one can also calibrate a few major
indices with a wide spectral baseline, like for example the {\rm (V\--K)} colour,
which turn out to be very sensitive to the stellar temperature. 
In this framework, F00,   
\citet[][\--hereafter V04]{io} and \citet[][\--hereafter S04]{BB}
presented near\--IR CMDs of a total sample of 16 GGCs (10 in F00, 5 in V04
and 1 in S04) which have been used to calibrate several
observables describing the RGB physical and chemical properties, and to
detect the major RGB evolutionary features (i.e the Bump and the Tip).
%
\begin{figure*}
\includegraphics[width=18cm]{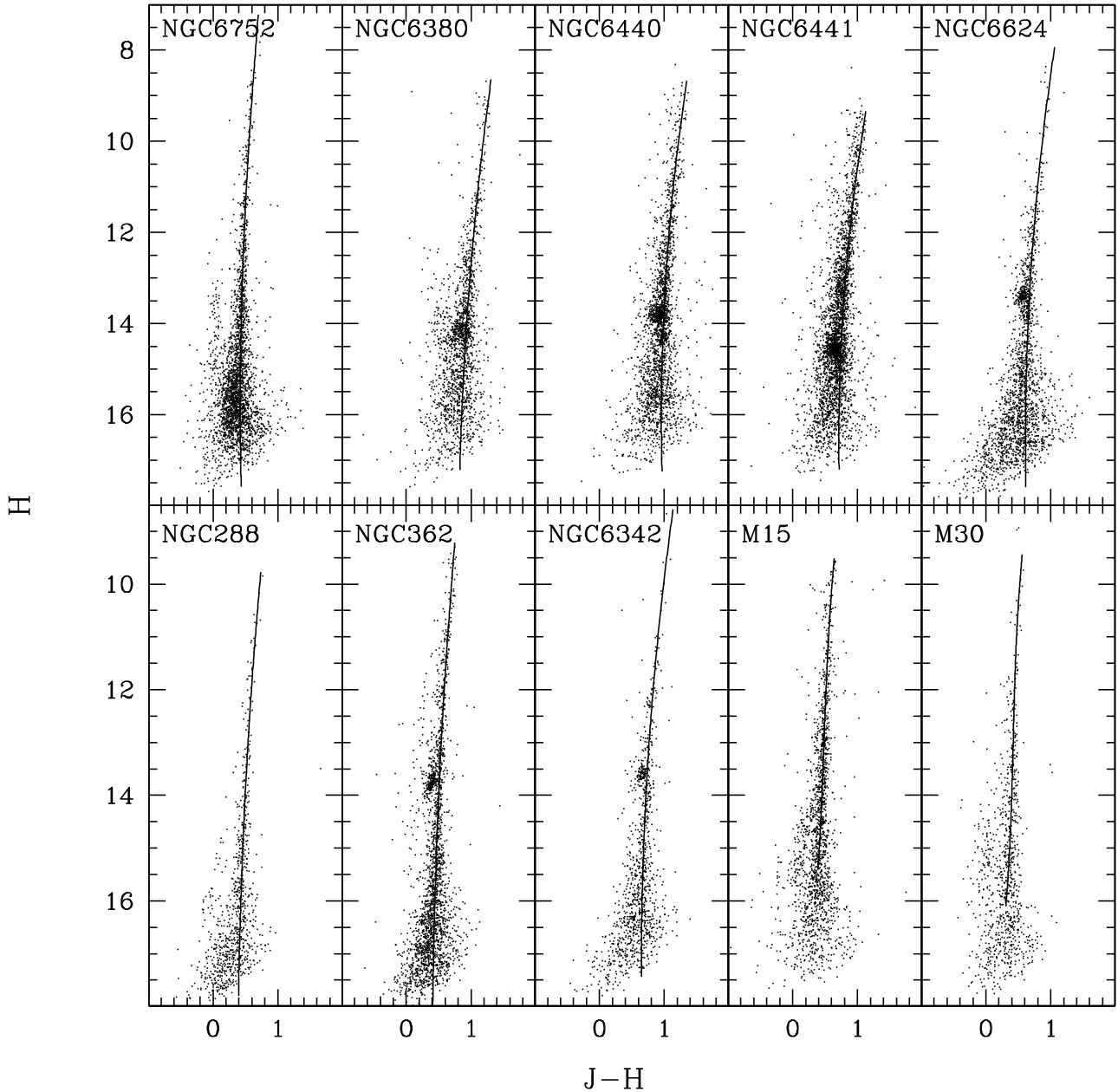}
\caption{H, J\--H colour\--magnitude diagrams for the 10 GGCs in our database.
The thick line in each panel indicates the RGB fiducial ridge line.} 
\label{hjh}
\end{figure*}

\begin{figure*}
\includegraphics[width=18cm]{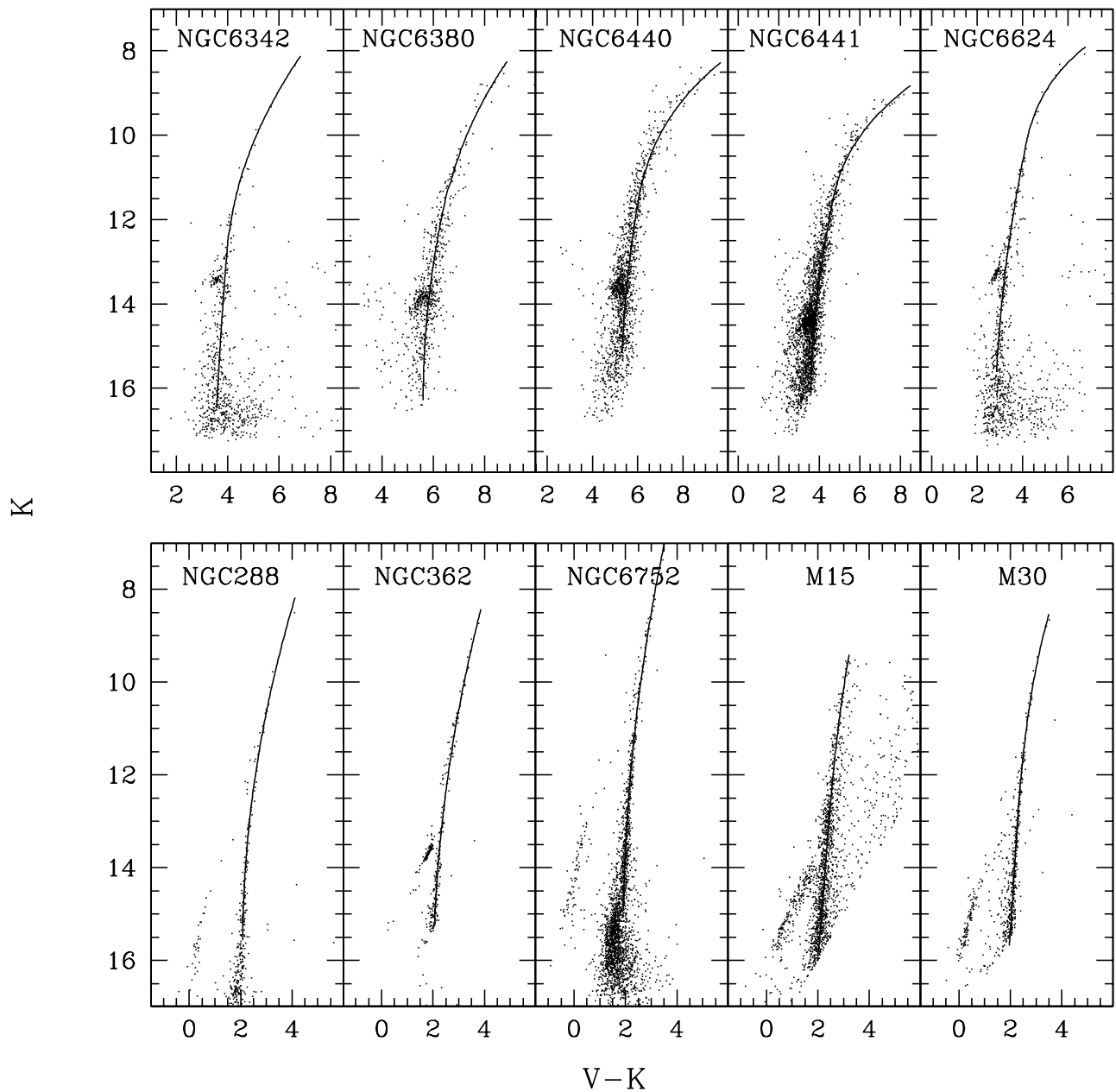}
\caption{K, V\--K colour\--magnitude diagrams for the 10 GGCs in our database.
The thick line in each panel indicates the RGB fiducial ridge line.} 
\label{kvk}
\end{figure*}
In this paper we present an addictional sample of 10 clusters belonging to
different Galactic populations: five clusters (namely NGC~288, NGC~362, 
NGC~6752, M~15 and M~30) belong to the Halo and five (namely NGC~6342, NGC~6380,
NGC~6441, NGC~6440 and NGC~6624) belong to the Bulge.
By combining the data set presented here and the data by F00, V04 and S04 
we have now available a homogeneous near\--IR database of 24 GGCs 
distributed over a wide metallicity range, 
$-2.12{\leq}{\rm [Fe/H]}{\leq}-0.49$.
In this first paper we presented the new data set and the calibration of the 
various RGB photometric parameters (colours at fixed magnitudes, magnitudes at 
fixed colours, slope) as a function of the cluster metallicity.
This work represent an update of the calibrations presented by F00,
based on a significative larger sample (especially in the high metallicity
domain).
Moreover, since H\--band observations were also availables we derive new 
calibrations of the RGB photometric indices in this band as well,
in order to have a more complete set of metallicity tracers in the
near\--IR bands.

A forthcoming paper (Valenti, Ferraro \& Origlia 2004, in preparation) will be
devoted to discuss the major evolutionary features (bump and tip)
and their calibration as a function of the metallicity. A third paper
(Ferraro et al. 2004, in preparation) will deal with the transformation
to the theoretical plane and the definition of useful relation to
empirically calibrate the mixing\--length parameter of theoretical models.

The observations and data reduction are presented in \S 2, while \S 3 describes
the properties of the observed CMDs. \S 4 is devoted to derive the mean RGB
features from the CMDs and the comparison with the
previous works. Finally, our conclusions are
summarized in \S 5.


\begin{table}
\begin{center}
\caption[]{\hspace{2.5cm}
{The observed sample}}
\label{sample}
\begin{tabular}{lc r}
\hline\hline
\\
Name&$[Fe/H]_{CG97}$&Optical \\
  &           &  Photometry\\
\hline
{\it Halo Clusters} &  \\
\\
M~15&-2.12&\citet{vM15}\\
M~30&-1.91&\citet{vM30}\\
NGC~6752&-1.42&\citet{F03}\\
NGC~362&-1.15&\citet{Bell01}\\
NGC~288&-1.07&\citet{Bell01}\\
\\
47~Tuc&-0.70&\citet{paolo95}\\
\\
\hline
{\it Bulge Clusters} & & \\
\\
NGC~6380&-0.87&\citet{v6380}\\
NGC~6342&-0.71&\citet{ptt02}\\
NGC~6441&-0.68&\citet{ptt02}\\
NGC~6624&-0.63&\citet{ptt02}\\
NGC~6440&-0.49&\citet{v6440}\\
\\
\hline
\end{tabular}
\end{center}
\end{table}
%
\section{Observations and data reduction}
A set of J, H and K images were secured at ESO, La Silla in August 1997, using
the ESO\--MPI 2.2m telescope equipped with the near\--IR camera IRAC\--2
\citep{irac2} based on a NICMOS\--3 $256{\times}256$ array detector. The central
$4'{\times}4'$ region of ten GGCs, 
namely NGC~288, NGC~362, NGC~6752, M~15, M~30, NGC~6342, NGC~6380, NGC~6440,
NGC~6441 and NGC~6624, were mapped
by using two different magnification: $0.28"/px$
 for the most crowed central field and $0.51 "/px$ for the four fields
centred at ${\sim}1'$ north\--east, north\--west, south\--east and south\--west 
of the cluster centre. 
An additional cluster, 47~Tuc, was also observed, but only
in the H band. Table~\ref{sample} lists the observed clusters and their 
metallicity in the \citet{CG97} scale.

During the four observing nights the average seeing was 1"\--1.2".
Each J, H and K image was the resulting average of 60 exposures of 1\--s
detector integration time (DIT)
and was sky\--subtract and flat\--field\--corrected. The sky field was located
several arcmin away from the cluster centre. More details on the pre\--reduction
procedure can be found in \citet{F94} and \citet{paolo95}.
The Point Spread Function (PSF) fitting procedure was performed
independently on each J, H and K image by using the ALLSTAR routine 
\citep{SH88} of the reduction package DAOPHOTII \citep{dao94}. A catalog
listing the instrumental J, H and K magnitudes for all the stars identified
in each field has been obtained by cross\--correlating the single band 
catalogs. All stars measured in at least two bands have been included in the
final catalog.
Since the observations were performed under not perfect photometric conditions,
we transformed the instrumental magnitudes into the
{\it Two Micron All Sky Survey} (2MASS) photometric system
{\footnote {In doing this we used the Second Incremental Release Point Source
Catalog of 2MASS}}. The large number of stars (typically a few hundreds) 
in the overlapping area between our observation and 2MASS survey
were used to derive the calibration to the 2MASS photometric system; only 
zero\--order polynomial relations, without colour terms, have been used.
 
Since M~15 and M~30 were observed also by F00, their photometric catalogs 
were combined with ours in order to reduce the photometric uncertainties. 
First, the catalog of M~15 and M~30 by F00 were transformed in the 2MASS 
photometric system by using the empirical transformations found by V04, 
then for each cluster we derived a unique catalog by averaging the 
multiple measurements.

An overall uncertainty of ${\pm}0.05$ mag in the zero point
calibration in all the three bands, has been estimated.
Fig.~\ref{hjh} and \ref{kvk}
show the H, {\rm J\--H} and K, {\rm V\--K} CMDs, respectively, for the
 observed clusters in
the 2MASS system.{\footnote {The observed cluster catalogs, in the 2MASS
photometric system are availables in the electronic form.}}  
%
\section{Colour Magnitude Diagrams}
More than $16,000$ and $9,000$ stars are plotted in the {\rm
(H,J\--H)} and {\rm (K,V\--K)} CMDs shown in Figs.~\ref{hjh} and \ref{kvk},
respectively. 
The references for the optical data used in this work are listed in 
Table~\ref{sample}. The main characteristic of the CMDs
are schematically summarized as follows:\\
{\it i)} The RGB is quite well populated in all the program clusters, even in
the brightest magnitude bin, and allows us a clean definition of the
mean ridge line, up to the end of the RGB.\\
{\it ii)} The observations are deep enough to
detect the base of the RGB at ${\Delta}K{\sim}{\Delta}H{\sim}7\--8$ 
mag fainter than the RGB tip, and ${\sim}$3\--4 mag below the Horizontal 
Branch (HB).\\
{\it iii)} In the combined CMDs the HB stars are clearly separable from the RGB
stars. For the Halo cluster sample, the HB has an almost vertical
structure in all the CMDs, as expected for a metal\--poor population.  
The Bulge clusters exhibit a red clumpy HB, which is typical of 
metal\--rich populations. In the case of NGC~6441, from the combined CMD it is
possible to clearly see the anomalous HB which exhibits 
both the typical features of metal\--poor and metal\--rich populations, 
a red clump and a populated blue branch 
\citep[see also][]{R97}.\\
\begin{table*}
\begin{center}
\caption{Reddening estimates for the program Bulge clusters.}
\label{red}
\begin{tabular}{lccccc}
\hline\hline
\\
Cluster&{\rm [Fe/H]$_{CG97}$}&{\rm E(B\--V)$_{Harris96}$}&
{\rm E(B\--V)$_{Schlegel98}$}&{\rm E(B\--V)$_{derived}$}&
{\rm E(B\--V)$_{adopted}$}\\
\\
\hline
\\
NGC~6342&-0.71&0.46&0.57&0.56&0.57\\
NGC~6380&-0.87&1.17&1.52&1.29&1.29\\
NGC~6441&-0.68&0.44&0.63&0.52&0.52\\
NGC~6624&-0.63&0.28&0.14&0.34&0.28\\
NGC~6440&-0.49&1.07&1.15&1.17&1.15\\
\\
\hline
\end{tabular}
\end{center}
\end{table*}
%
%
\subsection{Comparison with previous photometries}
Some of the program clusters, mainly those belonging to the Halo,
 have been the subject of several photometric and
spectroscopic observations in the optical bands. For example,
NGC~288 and NGC~362,
represent an HB {\it Second Parameter pair} \citep[see]
[ and references therein]{Bell01},
 and NGC~6441 has
been observed by several authors for its peculiar HB morphology
\citep[see][and references therein]{R97}. However,
only a few papers presented IR photometry for the clusters in our sample.
\citet{FPC83} reported J, H and K photometry of giants in NGC~288, NGC~362
and NGC~6752. A direct star\--to\--star comparison was not possible because
the authors did not published the coordinates of the observed stars; 
nevertheless their photometries nicely
overlap our IR\--CMDs with a minor offset of ${\approx}(0.03-0.05)$ mag.
The comparison of our K, J\--K CMD of NGC~288 with the
mean ridge line published by \citet{DH97} shows a good agreement. 
For M~15 and M~30
a comparison with previous photometries can be found in F00.\\
Conversely, for NGC~6440 and NGC~6624 a star\--to\--star comparison
between our data and the J, H and K photometry published by \citet{KF95} is
possible. They mapped a field of $2.5'{\times}2.5'$ centred ${\sim}1'$
north\--east from the centre in both clusters, using a $0.35"/px$ magnification. 
An offset of ${\approx}0.15$ mag
was found in all the three bands. Also \citet{minn95} presented 
IR\-- photometry of NGC~6440, but no online data are available, 
however their data agree with \citet{KF95}. Though the 2MASS photometric
system is different from that used by \citet{KF95} the measured offset 
seems too large to be due only to the different photometric systems.\\
IR photometric studies of NGC~6342 and NGC~6380 are not available in the
literature.
%
\section{The main RGB features}
The main aim of this series of papers is to present updated calibrations
of photometric RGB indices as a function of the metallicity, based on a complete
database collected by our group over the last 10 years, and presented in F00,
V04 and this paper.
In this section the RGB ridge lines and a few major
photometric indices, namely colours at fixed magnitudes and magnitudes
at fixed colours accordingly to the definitions by F00,
are derived from the CMDs shown in Fig.~\ref{hjh} and \ref{kvk}.
In order to properly combine this data set with those by F00 and V04,
we first need to make homogeneous the photometric systems. In particular, we
converted the photometry presented in F00 and V04 in the 2MASS  
system by using the relation found by V04. In the case of ${\omega}$~Cen,
the RGB ridge line was converted in the 2MASS photometric system
by using the offset found by S04 (${\Delta}$J=0.0 and ${\Delta}$K=-0.04).
After this transformation, a homogeneous data set of 24 clusters is
available. 
The RGB ridge lines and the photometric indices of the entire sample have
been newly determined.
Of corse all the known RGB variables lying in the region sampled
by our observations (see the case of 47~Tuc and NGC~6553 
in Figs. 1 and 2 of F00)
have been identified and removed from the RGB sample before
measuring any parameter.
\begin{figure}
\centering
\includegraphics[width=8.7cm]{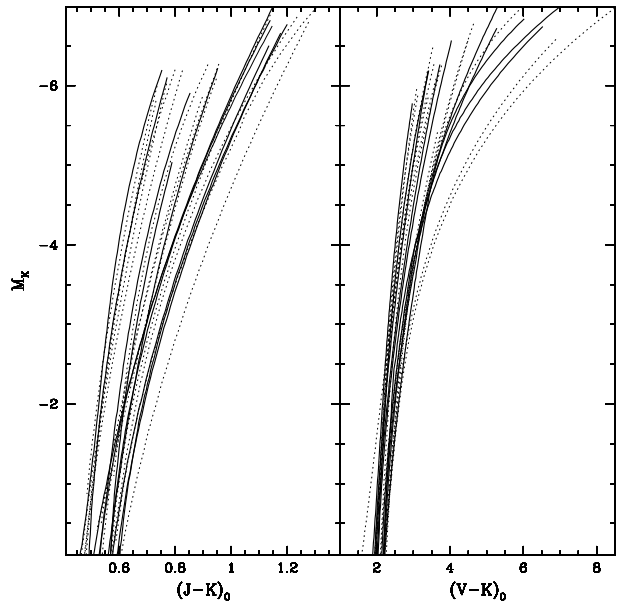} 
\caption{RGB fiducial ridge lines for the observed GGCs (solid lines) 
in the {\rm $M_K$,(J-K)$_0$} (left panel)
 and {\rm M$_K$,(V-K)$_0$} (right panel). The mean ridge lines for 
the clusters presented by F00, V04 and S04 (transformed in the 2MASS
photometric system) are plotted as dotted lines.}
\label{Klines}
\end{figure}

\begin{figure}
\centering
\includegraphics[width=8.7cm]{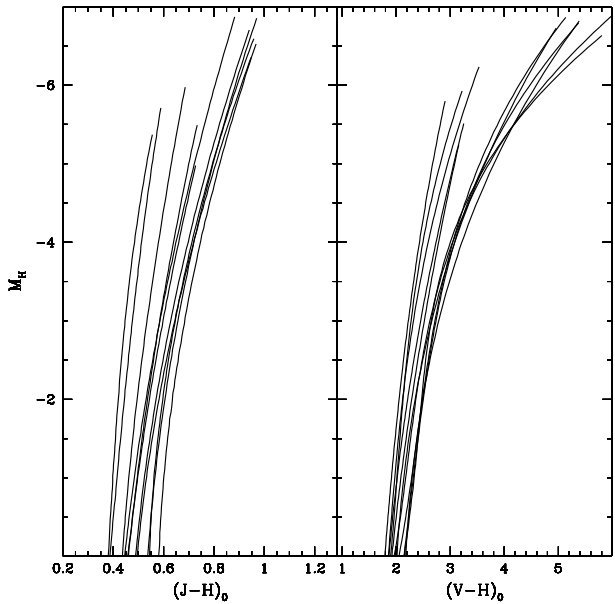} 
\caption{RGB fiducial ridge lines for the 10 observed GGCs and for 47~Tuc, 
in the {\rm $M_H$,(J-H)$_0$} (left panel)
 and {\rm M$_H$,(V-H)$_0$} (right panel).}
\label{Hlines}
\end{figure}
%
%
\subsection{The RGB fiducial ridge lines}
Since the procedure to obtain the RGB fiducial ridge lines for the 
observed clusters has been fully described in F00 and V04, 
it will not be repeated here. The ridge lines for the 10 clusters
presented here are overplotted to the
{\rm (H,J\--H)} and {\rm (K,V\--K)}
 CMDs shown in Fig.~\ref{hjh} and \ref{kvk}, respectively.
%
{\small 
\begin{table}
\begin{center}
\caption[]{Adopted parameters for the observed GGCs.}  
\label{newpar}
\begin{tabular}{lcccc}
\hline\hline
\\
Name & $[Fe/H]_{CG97}$ & $[M/H]$ & $E(B-V)$ & $(m-M)_0$ \\
\\
\hline
\\
$M~15^*$&-2.12& -1.91& 0.09& 15.15\\
$M~30^*$&-1.91& -1.71& 0.03& 14.71\\
$NGC~6752^*$&-1.42& -1.21& 0.04& 13.18\\
$NGC~362^*$&-1.15& -0.99& 0.05& 14.68\\
$NGC~288^*$&-1.07& -0.85& 0.03& 14.73\\
NGC~6380&-0.87& -0.68& 1.29& 14.81\\
NGC~6342&-0.71& -0.53& 0.57& 14.63\\
NGC~6624&-0.63& -0.48& 0.28& 14.63\\
NGC~6441&-0.68& -0.52& 0.52& 15.65\\
NGC~6440&-0.49& -0.40& 1.15& 14.58\\
$47~Tuc^*$&-0.70& -0.59& 0.04& 13.32\\
\\
\hline
\multicolumn{5}{l}{$^*$ For these clusters the 
estimates listed in Table 2 of F99 have}\\ 
\multicolumn{5}{l}{been used.}
\end{tabular}
\end{center}
\end{table}
}
%

\subsection{Reddening and distance modulus}
In order to transform the mean ridge lines into the absolute plane 
it is necessary
to adopt a distance scale and a reddening correction. The definition of the most
suitable distance scale for GGCs is still very controversial (see F99 and
references therein). In the present study, the distance scale established by
F99 was adopted. Nevertheless, in the F99 clusters
list (see their Table~2) only the Halo clusters sample are considered. For the
Bulge clusters we derived an independent distance modulus from the IR 
photometry presented here. In doing this, we compared the IR and
combined CMDs of the Bulge clusters with those of a 
{\it reference} cluster. This method allows, in principle, to derive 
simultaneously distance modulus and reddening estimates.
In fact, the needed colour and magnitude shifts to
overlap the CMDs of two clusters of comparable age and metallicity,
are a function of the reddening and distance differences, respectively.
Since several works on dating the Bulge GCs have showed that Halo and Bulge GCs
have comparable age \citep[see i.e][]{yaz03,Heas00,feljoh02,Ort01}, and
since our Bulge cluster sample has a metallicity comparable to that of
47~Tuc (within 0.2 dex, see Table\ref{sample}), we decided to adopt 47~Tuc
as a reference cluster. 
Moreover, the reddening, the metallicity and the distance 
of 47~Tuc are reasonably known, being one of the most studied GGC since 
many decades.
As can be seen from Table~\ref{red}, also the reddening determination 
of the Bulge clusters is quite uncertain (compare the values listed by 
\citet{redH96} with the most recent determination by \citet{redS98}).
Of course a different assumption on the reddening significantly affects
the position of the RGB in the absolute plane and the determination of the true
distance modulus. For this reason we used the differential analysis described 
above, in order to derive an independent reddening estimate and to decide 
the most appropriate reddening for each Bulge cluster in our sample. 
Of course, the position of the RGB in the CMD is a sensitive function of the 
metallicity, for this reason the differential method should be applied to 
clusters with similar metallicity.
From the relations found by F00 we estimate that a difference of 
${\approx}0.2$ dex in metallicity would produce a difference of ${\approx}0.04$
in the (J\--K) colour and ${\approx}$ 0.1 in (V\--K).\\
As can be seen from Table~\ref{sample}, three Bulge clusters in our sample
(namely, NGC~6342, NGC~6624 and NGC~6441) have a metallicity (in the 
CG97 scale) comparable to 47~Tuc (within 0.1 dex). NGC~6380 has a
nominal metallicity 0.2 dex lower than 47~Tuc, but the well defined HB clump
and the RGB shape suggest a higher metallicity, for this cluster. 
Previous papers \citep[e.g.][]{v6380} already suggested for NGC~6380 a 
metallicity between 47~Tuc and NGC~6553.
Finally, NGC~6440 is ${\approx}0.2$ dex more metal\--rich than the reference
cluster. 
We applied the differential method to the Bulge clusters in our sample, and
the shifts in colours in different planes (i.e. ${\delta}$(J\--H),
${\delta}$(J\--K), ${\delta}$(V\--J), ${\delta}$(V\--H),${\delta}$(V\--K)) have
been computed. Then, by adopting extinction coefficient for the V, J, H and K
band listed by \citet{SavMat} ({\rm A$_V/E(B\--V)=3.1$, A$_J/E(B\--V)=0.87$,
A$_H/E(B\--V)=0.54$ and A$_K/E(B\--V)=0.38$)we derived the average value 
for the reddening. The result of this procedure is shown in Table~\ref{red}.
As can be seen the value found by our procedure is similar to that found by
\citet{redS98} for NGC~6440 and NGC~6342, while it is more similar to the
\citet{redH96} value for NGC~6624. For two clusters in our sample, namely
NGC~6380 and NGC~6441, the reddening obtained by our procedure is significantly
different (and intermediate) from both the literature values. For these two
clusters we will adopt our reddening value. However, to be 
conservative, these two clusters are not considered in deriving the relations
between the position in colour of the RGB and the clusters metallicity
(in different planes).
By assuming the reddening listed in column [6] of Table~\ref{red}
we derived the distance modulus
by comparison with 47~Tuc. The HB clump has been chosen as a reference
sequence.

The adopted method can be summarized as follow:\\
{\it i)} The LFs in the IR passbands have been constructed to identifying the HB
peak, which it is been used as HB level.\\
{\it ii)} By using the LFs we measured the differences between the 
47~Tuc HB level and those of the Bulge clusters; the derived values have been
adopted to shift the clusters CMD on the reference one.\\
{\it iii)} Finally, the differences in magnitudes measured in the various
bands have been corrected for reddening (by using the relations
quoted above) and the true distance modulus has been obtained.\\
It is worth noting that in applying this method, all the available photometric 
bands were used in order to get a more careful estimate. 
Table~\ref{newpar} lists the adopted distance 
modulus for all the program clusters. 

Fig.~\ref{Klines} shows the observed RGB fiducial ridge lines in the absolute
{\rm M$_K$}, {\rm (J-K)$_0$} and {\rm M$_K$}, {\rm (V-K)$_0$}
 planes for the entire database of 24 GGCs (the 10 clusters presented here
are plotted as solid lines).
As expected, the mean ridge lines
of our 5 intermediate\--low metallicity clusters lie in the bluer region of the
diagrams, while in the redder part we find those of high\--metallicity clusters
of the Bulge. A similar behavior can be seen in Fig.~\ref{Hlines},
 which shows the RGB ridge lines in the absolute
{\rm M$_H$}, {\rm (J-H)$_0$} and {\rm M$_H$}, {\rm (V-H)$_0$} planes.
In the {\rm M$_H$}, {\rm (V-H)$_0$}
plane, the two different groups are more clearly distinguished. 
The Halo cluster RGB
lines are bluer and less curved than the RGB lines of the more metal\--rich
Bulge clusters.
\begin{figure}
\centering
\includegraphics[width=8.7cm]{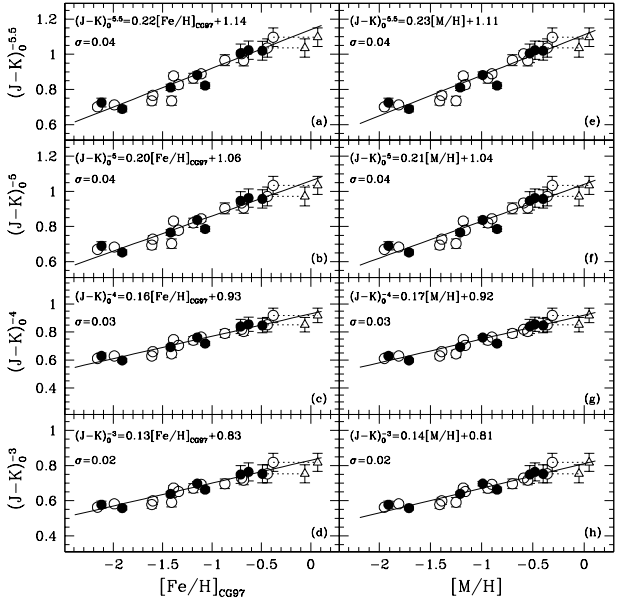} 
\caption{RGB mean {\rm (J\--K)$_0$} colour at fixed ({\rm
M$_K=-5.5, -5, -4, -3$}) magnitudes as a function of the CG97 metallicity scale
(left panels) and of the global metallicity (right panels).
Filled circles: the 10 clusters observed here. Empty circles: the F00, V04 and
S04 samples. 
The empty 
triangles refer to NGC~6553 and NGC~6528 adopting the \citet{eugi01} 
metallicity
estimates. The solid lines are best\--fitting relations.}
\label{coljk}
\end{figure}

\begin{figure}
\centering
\includegraphics[width=8.7cm]{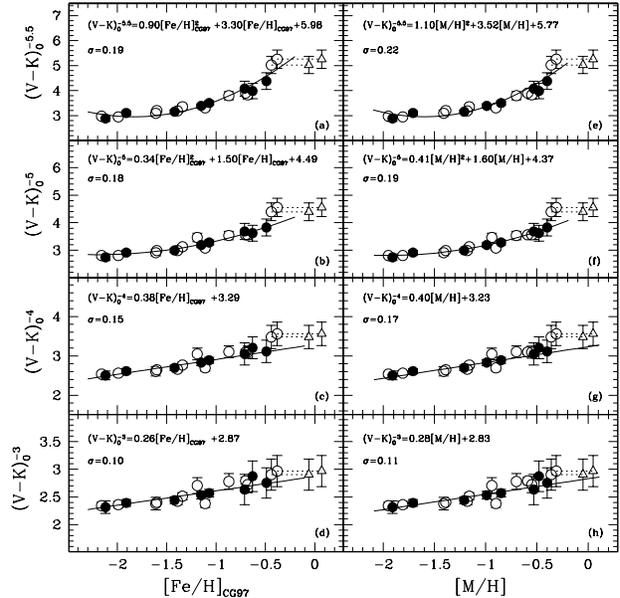} 
\caption{The same as Fig.~\ref{coljk}, but for {\rm (V\--K)$_0$} colours.}
\label{colvk}
\end{figure}
\subsection{The RGB location in Colour and in Magnitude}
As already discussed in detail by F00, to properly characterize the overall
behavior of the RGB as a function of the cluster metallicity, a set of
photometric indices are needed (see \S 4).
In fact, at fixed colours the corresponding
magnitudes mark different RGB regions, depending on the clusters metallicity.
Several parameters describing the RGB location in colour and in magnitude have
been suggested by many authors (see F00 and references therein). Nevertheless, 
to get a complete description of the RGB photometric properties, in
the present study we use the new parameters defined by
F00, namely the {\rm (J\--K)$_0$} and {\rm (V\--K)$_0$} colours at different
absolute magnitudes {\rm M$_K=-3, -4, -5, -5.5$}, and the K absolute magnitude 
at fixed {\rm (J\--K)$_0$} and {\rm (V\--K)$_0$} colours, respectively.
The derived {\rm (J\--K)$_0$} and {\rm (V\--K)$_0$} RGB colours for the program 
clusters are listed in Table~\ref{jk} and \ref{vk}, respectively. 
In both tables, the
measurements by F00 and V04, converted in the 2MASS photometric system, are also
reported. The colours at fixed magnitudes for all the clusters in the database
have been calibrated as a function of:
{\it i)} the metallicity in the CG97 scale, and 
{\it ii)} the global metallicity {\rm ($[M/H]$)} defined and computed in F99,
which takes into account the contribution of the ${\alpha}$\--elements in the
definition of the global metallicity of the cluster.
The metallicity in the CG97 scale for the program clusters
has been computed from the Zinn (1985) scale by 
using equation [7] of CG97, following the prescriptions by F99.
The typical uncertainty on the derived metallicities can be conservatively
assumed to be 0.2 dex; however, for clusters having direct
CG97 measurements the error is significantly lower, $<$0.1 dex, 
(see Table [8] of CG97).
 
The calibration relations of the RGB photometric indices as function of the
cluster metallicity in both the adopted scales are listed in the Appendix. 

The case of NGC~6553 and NGC~6528 (the two clusters which represent the
metal\--rich extreme of our entire database) deserves a few additional comments.
The metallicity of these two clusters has been, in fact, largely debated
in the literature.
By simply considering the most recent determinations based on high resolution 
spectroscopy, values ranging from -0.3 up to about solar 
\citep{eugi01,livia02,mele03} have been proposed.
To be homogeneous with other clusters, for NGC~6553 and NGC~6528 in the
following calibrations we will adopt the CG97 values listed in Table\ref{coljk}.
Figs.~\ref{coljk} and \ref{colvk} show the {\rm (J\--K)$_0$ and
(V\--K)$_0$} colours as a function of both the CG97 and global metallicity
scales, for the entire sample of 24 clusters.
By using the full data set, updated calibrations have been derived and reported
in each panel and in the Appendix. 
As can be seen from Fig.~\ref{coljk} the RGB {\rm (J\--K)$_0$}
colours linearly scale with the metallicity. As expected from previous studies
\citep[see][ and F00]{cs95}
the fit slope increases progressively toward the RGB tip. The
derived slope values are consistent with those found by F00. Conversely,
in the {\rm (V\--K)$_0$} plane, the best\--fitting
solution deviates from a linear dependence at higher metallicity 
(see Fig.~\ref{colvk}, panels {\it a, b, e, f}) 
even if the \citet{eugi01} metallicity estimates
for the most metal\--rich clusters are adopted. 
As can be seen the RGB, particularly near the tip, rapidly becomes redder and
redder as the metallicity increases as shown 
by \cite{cs95} and successively confirmed by F00.\\
For NGC~6624, \citet{cs95} derived the {\rm (J\--K)$_0$ and (V\--K)$_0$} colours
at fixed absolute magnitude {\rm M$_K$=-4, -5}. Their estimates in the {\rm K,
(J\--K)} plane (see their Table 10) are systematically redder, by ${\sim}0.15$
with respect to our determinations.
This is due to different reddening and distance assumptions:  
when we apply their reddening and distance modulus values to our photometry,
the difference in the derived {\rm (J\--K)$_0$} colours is reduced to only
${\sim}0.03$ mag. In the {\rm K, (V\--K)} plane, a ${\sim}0.1$ mag difference
remains even when the same reddening and distance modulus are adopted. 
Conversely, a nice agreement in the derived 
{\rm (V\--K)$_0^{M_K=-5}$} colour
was found with the value published by \citet{KF95}.\\

\begin{figure}
\centering
\includegraphics[width=8.7cm]{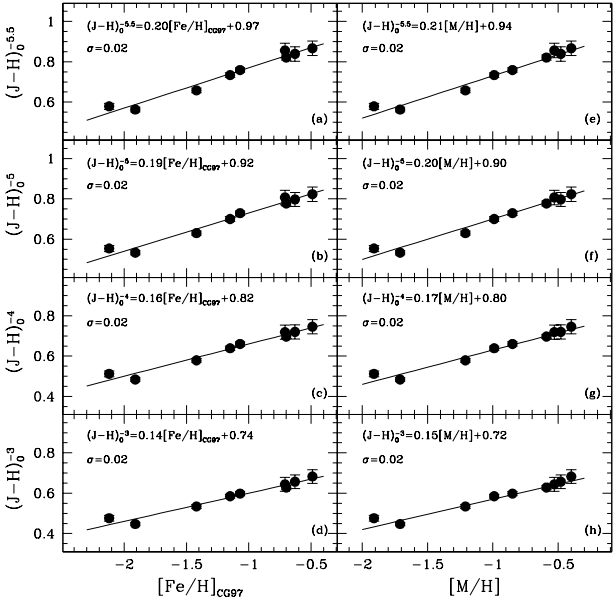} 
\caption{ RGB mean {\rm (J\--H)$_0$} colour at fixed ({\rm
M$_H=-5.5, -5, -4, -3$}) magnitudes as a function of the CG97 metallicity 
scale
(left panels) and of the global metallicity (right panels) for the observed
clusters. The solid lines are best\--fitting relations.}
\label{coljh}
\end{figure}

\begin{figure}
\centering
\includegraphics[width=8.7cm]{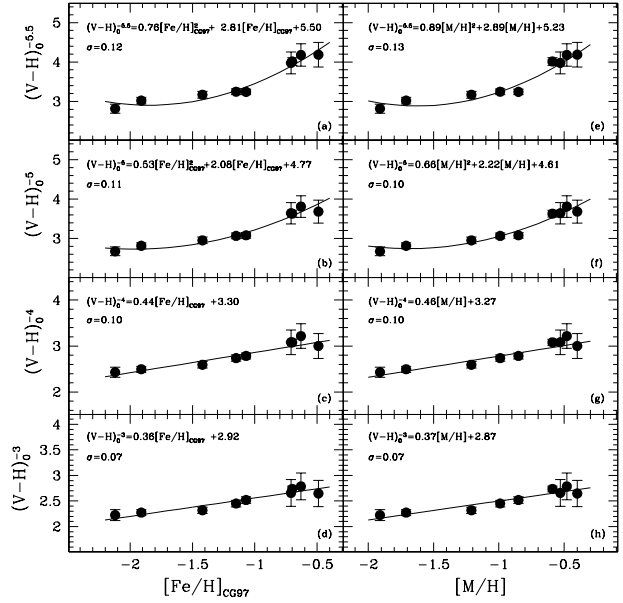} 
\caption{The same as Fig.~\ref{coljh}, but for {\rm (V\--H)$_0$} colours.}
\label{colvh}
\end{figure}

By using {\rm (J\--H)$_0$} and {\rm (V\--H)$_0$} colours at different absolute
magnitudes {\rm M$_H=(-3, -4, -5, -5.5)$}, new calibrations have been 
proposed in
the H band. The derived values for the program clusters are listed in
Table~\ref{jh} and \ref{vh}, while Figs.~\ref{coljh} and \ref{colvh} 
show the behavior of the {\rm (J\--H)$_0$ and (V\--H)$_0$} colours, respectively,
as a function of the cluster metallicity in both the adopted metallicity 
scales. The best fits to the data are shown in each panel and 
listed in the Appendix. 
As expected, the colours become redder with increasing
clusters metallicity in a linear way and independently from the height cut in
the {H, (J\--H)} plane, while at brighter magnitudes the {\rm (V\--H)$_0$} 
colour shows a quadratic metallicity dependence.\\

\begin{figure}
\centering
\includegraphics[width=8.7cm]{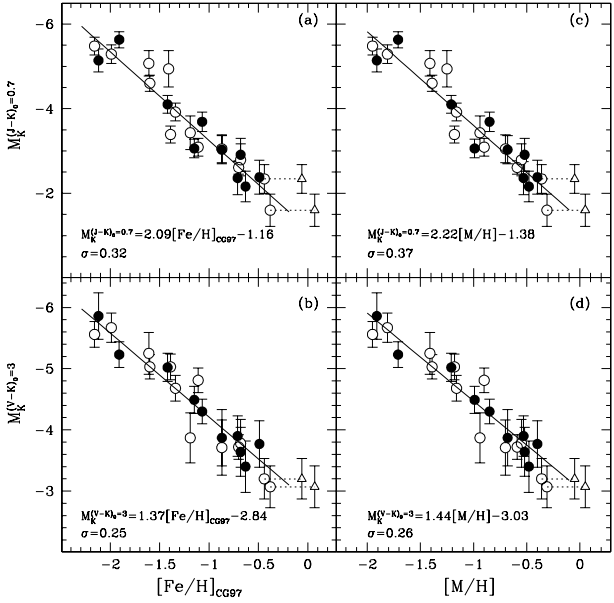}
\caption{Upper panels: {\rm M$_K$} at fixed {\rm (J\--K)$_0=0.7$} as a
function of the metallicity in the CG97 (a) and in the global (c) scale. 
Lower panels: {\rm M$_K$} at constant {\rm (V\--K)$_0=3$} as a function of the
CG97 (b) and global (d) metallicity. The filled circles refer to the present
sample, the empty circles mark the F00, V04 and S04 data and the empty 
triangles point NGC~6553 and NGC~6528 
adopting the \citet{eugi01} metallicity estimates. 
The solid lines are best\--fitting relations.}
\label{magK}
\end{figure} 

\begin{figure}
\centering
\includegraphics[width=8.7cm]{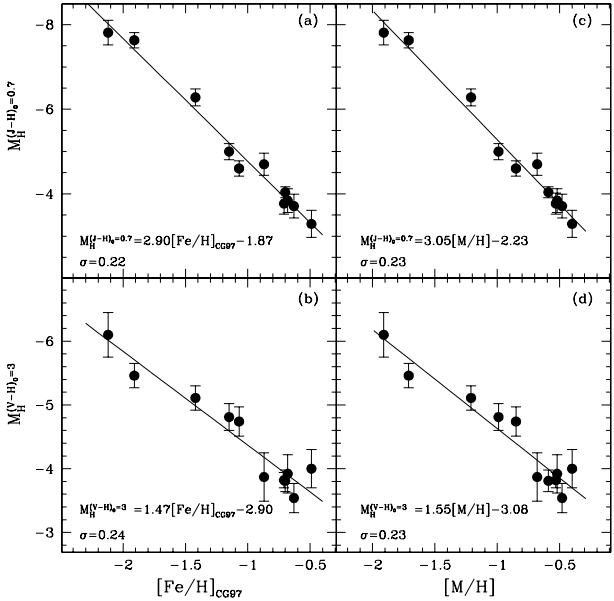}
\caption{The same as Fig.~\ref{magK}, but for M$_H$ magnitudes.}
\label{magH}
\end{figure}
%
Following \citet{FCP83} and F00 we also measured the K absolute magnitude 
at fixed {\rm (V\--K)$_0=3$ and (J\--K)$_0=0.7$} colours. 
In Fig.~\ref{magK} we show 
the dependence of these parameters on metallicity in both the adopted scales, 
for the entire sample. 
The best\--fitting relations are
also reported in each panel. Table~\ref{jk} and \ref{vk}
list the derived {\rm M$_K $}magnitudes at constant {\rm (J\--K)$_0$} and {\rm
(V\--K)$_0$} colours, respectively.
While the error associated to the determination of the colours at fixed
absolute magnitudes are mainly driven by the uncertainty on the distance
modulus, the accuracy on the derived absolute magnitude at fixed colours 
depends on both distance and reddening uncertainties with almost the same 
weight. In fact,
given the intrinsic steepness of the RGB, especially in the metal\--poor range, 
an error of a few hundredths of magnitude in the reddening correction easily
implies 0.15\--0.20 mag uncertainty in the derived {\rm M$_K$} absolute
magnitudes, depending on the height along the RGB (see Fig.~\ref{Klines}).\\
By using the same strategy we also derive the {\rm M$_H$} absolute magnitude at
fixed {\rm (J\--H)$_0=0.7$ and (V\--H)$_0=3$} colours, listed in 
Tables~\ref{jh} and \ref{vh} and plotted in Fig.~\ref{magH} as a function of 
the metallicity in both the adopted scales. 
The best\--fitting relations with the corresponding standard deviation are
reported in each panel and listed in the Appendix.
%
\begin{figure}
\centering
\includegraphics[width=8.7cm]{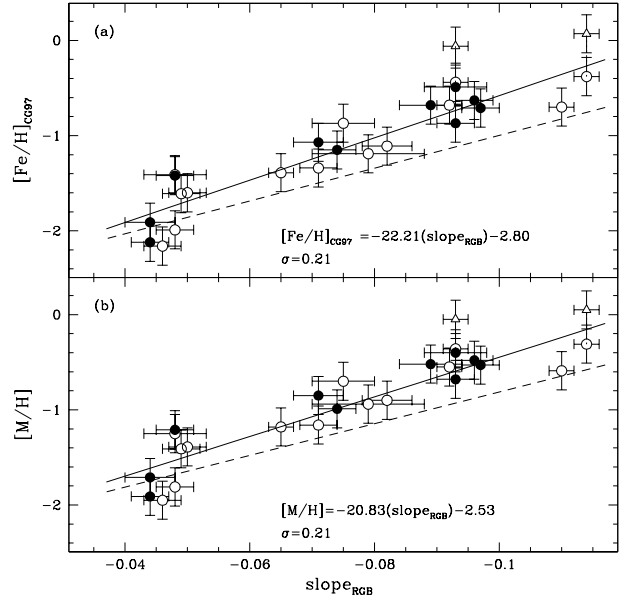}
\caption{Metallicity scale: [Fe/H]$_{CG97}$ (a), and [M/H] (b) as a
 function of the derived RGB slope for the selected 10 GCs (filled circles) 
and for the F00, V04 and S04 program clusters (empty circles). 
The solid lines are our best\--fitting 
relations, while the dashed lines are the relations found by \citet{IB02}.}
\label{slope}
\end{figure}
\subsection{The RGB slope}
An useful parameter to provide a photometric estimate of the cluster 
metallicity is the so\--called RGB slope. This parameter turns to be extremely
powerful since it is independent from reddening and distance.
Nevertheless, a careful estimate of the RGB slope is a complicated
task, even in the K, {\rm (J\--K)} plane, where the RGB is steeper than
in any other plane. 
As shown by \citet{K95,KF95}, a reasonable description of
the overall RGB morphology can be obtained by linearly fitting
the RGB in the range between 0.6 and 5.1 magnitudes brighter than the Zero
Age Horizontal Branch (ZAHB). However, in the case of low\--intermediate
metallicity clusters the accurate measurement of the location of the ZAHB in the
IR CMD is an almost impossible task, because the HB is not horizontal at all.
In order to apply an homogeneous procedure to the entire cluster sample, we 
fit the RGB in a magnitude
range between 0.5 and 5 magnitudes fainter than the brightest
star of each cluster after a previous decontamination
by the Asymptotic Giant Branch (AGB) and
field stars. In particular, in the case of the Bulge clusters, the 
level of field contamination
was estimated from the comparison with a {\it field}\--CMD obtained from the 
2MASS catalog for an equivalent area ($4{\arcmin}{\times}4{\arcmin}$) 
located at $10{\arcmin}$ from
the clusters center. On the basis of this comparison, a typical  
bulge contamination of 20\% was found in the RGB region. Then 
the estimated number of field stars has been randomly
removed from the cluster RGB sample, before determining the RGB slope.
The derived RGB slope values for the entire  
sample are listed in Table~\ref{jk}.  
Fig.~\ref{slope} shows the linear correlation of the
RGB slope with the metallicity (in both the adopted scales);
 the inferred relations, with the corresponding standard
deviations, are also reported in each panel.
As expected the RGB slope becomes progressively steeper
with decreasing metallicity, confirming the results found by
\citet{K95,KF95} and F00.
The considerable disagreement between our results and the
inferred relations found by \citet{IB02} (dashed lines in Fig.~\ref{slope}) 
in particular in the high metallicity range,
are mainly due to two different reasons: {\it i)} their sample of 22 GCs
includes only 3 clusters more metal\--rich than {\rm $[Fe/H]_{CG97}=-1$} 
and none more metallic than 47~Tuc, while our
best\--fitting relations are based on a global sample of 24 clusters, among them
7 more metal\--rich than 47~Tuc,
{\it ii)} the discrepancy in the estimate of the 47~Tuc RGB slope
($-0.110{\pm}0.002$, F00 and $-0.125{\pm}0.002$ \citet{IB02}). Indeed,
\citet{IB02} computed a weighed average relation which turned to be
significantly influenced by the value of 47~Tuc, being the cluster
with the most accurate determination.
%
%
\section{Summary and conclusions}
A new set of high\--quality IR CMDs for a sample of 10 GGCs
spanning a wide metallicity range have been presented. 
This database has been combined with the data set collected by our group
over the last 10 years (see F00, V04 and S04) and it has been 
used to measure a few
major observables describing the main photometric properties of the
RGB, namely: {\it i)} the location in colour and in magnitude, and
 {\it ii)} its slope. \\
The behaviour of these quantities as a function of the
clusters metallicity has been studied in both {\rm $[Fe/H]_{CG97}$} and [M/H]
metallicity scales. Since our database also
include observations in the H\--band, it has been used to derive for the
first time the calibrations in the H, J\--H and H, V\--H planes, as well.
All the relations are reported in the corresponding panels of
Figs.~\ref{coljk}\--\ref{slope} and in the Appendix, for more clarity.
%
\begin{table*}
\begin{center}
\caption[]{RGB location in colour (columns [4, 5, 6, 7]), in magnitude
(column [8]) in the K, J\--K plane and the RGB slope
for the observed GCs and for the F00 ,V04 and S04 samples.} 
\label{jk}
\begin{tabular}{lcccccccc}
\hline\hline
\\
Name & $[Fe/H]_{CG97}$ & $[M/H]$ & $(J-K)_0^{-5.5}$ & $(J-K)_0^{-5}$ &
$(J-K)_0^{-4}$ &$(J-K)_0^{-3}$ & $M_K^{(J\--K)=0.7}$&RGB$_{Slope}$\\
\\
\hline
\\
M 15  &     -2.12 &  -1.91 & $0.725{\pm}0.025$ &  $0.690{\pm}0.023$ &
  $0.629{\pm}0.022$ &  $0.577{\pm}0.020$& $-5.14{\pm}0.27$&$-0.044{\pm}0.003$ \\
M 30  &     -1.91 &  -1.71 & $0.689{\pm}0.016$ &  $0.653{\pm}0.014$ &
  $0.597{\pm}0.012$ &  $0.558{\pm}0.011$& $-5.63{\pm}0.19$&$-0.044{\pm}0.004$ \\ 
NGC 6752 &  -1.42 &  -1.21 & $0.811{\pm}0.018$ &  $0.766{\pm}0.016$ &
  $0.693{\pm}0.014$ &  $0.639{\pm}0.012$& $-4.10{\pm}0.21$&$-0.048{\pm}0.003$ \\  
NGC 362  &  -1.15 &  -0.99 & $0.882{\pm}0.017$ &  $0.837{\pm}0.017$ &
  $0.761{\pm}0.014$ &  $0.697{\pm}0.013$& $-3.06{\pm}0.22$&$-0.074{\pm}0.003$ \\  
NGC 288 &   -1.07 &  -0.85 & $0.822{\pm}0.015$ &  $0.786{\pm}0.015$ &
  $0.718{\pm}0.014$ &  $0.663{\pm}0.013$& $-3.69{\pm}0.23$&$-0.071{\pm}0.004$ \\  
NGC 6380 &  -0.87 &  -0.68 &0.954${\pm}0.052$&0.895${\pm}0.052$&
0.789${\pm}0.051$&0.697${\pm}0.050$& $-3.03{\pm}0.33$&$-0.094{\pm}0.003$ \\  
NGC 6342 &  -0.71 &  -0.53 & $1.005{\pm}0.053$ &  $0.946{\pm}0.052$ &
  $0.840{\pm}0.051$ &  $0.749{\pm}0.051$& $-2.36{\pm}0.39$&$-0.102{\pm}0.003$ \\  
NGC 6441 &  -0.68 &  -0.52 &0.958${\pm}0.053$&0.898${\pm}0.052$&
0.792${\pm}0.051$&0.707${\pm}0.050$& $-2.91{\pm}0.39$&$-0.092{\pm}0.005$ \\  
NGC 6624 &  -0.63 &  -0.48 & $1.023{\pm}0.052$ &  $0.962{\pm}0.052$ &
  $0.855{\pm}0.051$ &  $0.764{\pm}0.051$& $-2.16{\pm}0.36$&$-0.095{\pm}0.003$ \\  
NGC 6440 &  -0.49 &  -0.40 & $1.020{\pm}0.053$ &  $0.957{\pm}0.052$ &
  $0.847{\pm}0.051$ &  $0.753{\pm}0.051$& $-2.38{\pm}0.40$&$-0.093{\pm}0.005$ \\  
\\
M 68  &     -1.99 &  -1.81 & $0.712{\pm}0.013$ &  $0.683{\pm}0.013$ &
  $0.629{\pm}0.012$ &  $0.582{\pm}0.012$& $-5.29{\pm}0.22$&$-0.048{\pm}0.003$ \\
M 55  &     -1.61 &  -1.41 & $0.735{\pm}0.023$ &  $0.694{\pm}0.023$ &
  $0.629{\pm}0.021$ &  $0.578{\pm}0.021$& $-5.07{\pm}0.30$&$-0.049{\pm}0.003$ \\
M 4   &     -1.19 &  -0.94 & $0.864{\pm}0.028$ &  $0.821{\pm}0.027$ &
  $0.741{\pm}0.027$ &  $0.671{\pm}0.026$& $-3.43{\pm}0.40$&$-0.079{\pm}0.009$ \\
M 107 &     -0.87 &  -0.70 & $0.966{\pm}0.031$ &  $0.903{\pm}0.031$ &
  $0.790{\pm}0.029$ &  $0.696{\pm}0.027$& $-3.05{\pm}0.33$&$-0.075{\pm}0.005$ \\
47 Tuc &    -0.70 &  -0.59 & $1.003{\pm}0.018$ &  $0.934{\pm}0.016$ &
  $0.819{\pm}0.014$ &  $0.729{\pm}0.012$& $-2.61{\pm}0.16$&$-0.110{\pm}0.002$ \\
M 69  &     -0.68 &  -0.55 & $0.964{\pm}0.031$ &  $0.906{\pm}0.030$ &
  $0.804{\pm}0.028$ &  $0.717{\pm}0.027$& $-2.79{\pm}0.38$&$-0.092{\pm}0.002$ \\
NGC 6553 &     -0.44 &  -0.36 & $1.036{\pm}0.052$ &  $0.971{\pm}0.053$ &
  $0.852{\pm}0.052$ &  $0.753{\pm}0.051$& $-2.34{\pm}0.34$&$-0.092{\pm}0.002$ \\
NGC 6528 &     -0.38 &  -0.31 & $1.097{\pm}0.053$ &  $1.034{\pm}0.052$ &
  $0.919{\pm}0.052$ &  $0.818{\pm}0.051$& $-1.60{\pm}0.38$&$-0.114{\pm}0.002$ \\
\\
M 92  &     -2.16 &  -1.95 & $0.701{\pm}0.014$ &  $0.670{\pm}0.013$ &
  $0.611{\pm}0.013$ &  $0.563{\pm}0.012$& $-5.48{\pm}0.21$&$-0.046{\pm}0.003$ \\
M 10  &     -1.41 &  -1.25 & $0.735{\pm}0.026$ &  $0.703{\pm}0.026$ &
  $0.644{\pm}0.026$ &  $0.591{\pm}0.026$& $-4.94{\pm}0.43$&$-0.048{\pm}0.005$ \\
M 13  &     -1.39 &  -1.18 & $0.877{\pm}0.018$ &  $0.831{\pm}0.017$ &
  $0.746{\pm}0.015$ &  $0.672{\pm}0.014$& $-3.39{\pm}0.20$&$-0.065{\pm}0.002$ \\
M 3   &     -1.34 &  -1.16 & $0.827{\pm}0.019$ &  $0.779{\pm}0.016$ &
  $0.705{\pm}0.013$ &  $0.652{\pm}0.012$& $-3.92{\pm}0.21$&$-0.071{\pm}0.003$ \\
M 5   &     -1.11 &  -0.90 & $0.889{\pm}0.017$ &  $0.844{\pm}0.016$ &
  $0.764{\pm}0.015$ &  $0.693{\pm}0.014$& $-3.09{\pm}0.21$&$-0.082{\pm}0.004$ \\
\\
${\omega}$~Cen & -1.60 & -1.39 & $0.766{\pm}0.020$&$0.728{\pm}0.020$&$0.660{\pm}0.020$& 
$0.599{\pm}0.020$& $-4.602{\pm}0.19$& $-0.050{\pm}0.003$\\
\\
\hline
\end{tabular}
\end{center}
\end{table*}

\begin{table*}
\begin{center}
\caption[]{RGB {\rm (V\--K)$_0$} colours at fixed magnitudes {\rm (M$_K=-5.5,
-5, -4, -3$)} and K absolute magnitude at constant {\rm (V\--K)$_0$} colour
for the observed GCs and for the F00, V04 and S04 samples.}
\label{vk}
\begin{tabular}{lccccccc}
\hline\hline
\\
Name & $[Fe/H]_{CG97}$ & $[M/H]$ & $(V-K)_0^{-5.5}$ & $(V-K)_0^{-5}$ &
$(V-K)_0^{-4}$ &$(V-K)_0^{-3}$& $M_K^{(V\--K)_0=3}$\\
\\
\hline
\\
M 15   & -2.12&-1.91& $ 2.886{\pm}0.118$&$2.743{\pm}0.116$&
   $2.505{\pm}0.113$&$2.315{\pm}0.112$&  $-5.86{\pm}0.38$\\  
M 30   & -1.91&-1.71& $ 3.106{\pm}0.083$&$2.914{\pm}0.077$&
   $2.611{\pm}0.067$&$2.392{\pm}0.062$&  $-5.23{\pm}0.21$\\  
NGC 6752  & -1.42&-1.21& $ 3.157{\pm}0.074$&$2.993{\pm}0.072$&
   $2.696{\pm}0.068$&$2.441{\pm}0.064$&  $-5.02{\pm}0.23$\\  
NGC 362   & -1.15&-0.99& $ 3.389{\pm}0.083$&$3.189{\pm}0.080$&
   $2.831{\pm}0.075$&$2.532{\pm}0.068$&  $-4.49{\pm}0.22$\\  
NGC 288   & -1.07&-0.85& $ 3.504{\pm}0.089$&$3.280{\pm}0.085$&
   $2.889{\pm}0.076$&$2.569{\pm}0.069$&  $-4.30{\pm}0.20$\\  
NGC 6380  & -0.87&-0.68& 3.938${\pm}0.294$& 3.601${\pm}0.288$&
3.051${\pm}0.291$ & 2.703${\pm}0.276$ &  $-3.87{\pm}0.46$\\  
NGC 6342  & -0.71&-0.53& $ 4.078{\pm}0.301$&$3.681{\pm}0.295$&
   $3.049{\pm}0.284$&$2.635{\pm}0.276$&  $-3.90{\pm}0.33$\\  
NGC 6441  & -0.68&-0.52&4.167${\pm}0.331$ & 3.674${\pm}0.302$ &
3.132${\pm}0.278$ &2.770${\pm}0.277$&  $-3.64{\pm}0.40$\\  
NGC 6624  & -0.63&-0.48& $ 3.985{\pm}0.308$&$3.622{\pm}0.288$&
   $3.204{\pm}0.278$&$2.875{\pm}0.276$&  $-3.40{\pm}0.42$\\  
NGC 6440  & -0.49&-0.40& $ 4.380{\pm}0.337$&$3.827{\pm}0.311$&
   $3.113{\pm}0.284$&$2.754{\pm}0.275$&  $-3.77{\pm}0.38$\\  
\\
M 68   & -1.99&-1.81& $ 2.949{\pm}0.070$&$2.808{\pm}0.067$&
   $2.562{\pm}0.064$&$2.360{\pm}0.061$&  $-5.67{\pm}0.24$\\  
M 55   & -1.61&-1.41& $ 3.094{\pm}0.124$&$2.910{\pm}0.121$&
   $2.609{\pm}0.116$&$2.379{\pm}0.113$&  $-5.25{\pm}0.34$\\  
M 4    & -1.19&-0.94& \------&$3.464{\pm}0.152$&
   $3.049{\pm}0.148$&$2.706{\pm}0.144$&  $-3.87{\pm}0.41$\\  
M 107  & -0.87&-0.70& $ 3.798{\pm}0.161$&$3.535{\pm}0.155$&
   $3.105{\pm}0.147$&$2.780{\pm}0.142$&  $-3.71{\pm}0.45$\\  
47 Tuc & -0.70&-0.59& $ 3.900{\pm}0.099$&$3.559{\pm}0.081$&
   $3.098{\pm}0.066$&$2.792{\pm}0.060$&  $-3.72{\pm}0.20$\\  
M 69   & -0.68&-0.55& $ 3.830{\pm}0.161$&$3.559{\pm}0.157$&
   $3.094{\pm}0.150$&$2.723{\pm}0.145$&  $-3.86{\pm}0.33$\\  
NGC 6553  & -0.44&-0.36& $ 5.023{\pm}0.346$&$4.396{\pm}0.323$&
   $3.480{\pm}0.294$&$2.904{\pm}0.281$&  $-3.20{\pm}0.33$\\  
NGC 6528  & -0.38&-0.31& $ 5.255{\pm}0.365$&$4.553{\pm}0.334$&
   $3.561{\pm}0.298$&$2.968{\pm}0.281$&  $-3.07{\pm}0.34$\\
\\
M 92   & -2.16&-1.95& $ 2.978{\pm}0.078$&$2.808{\pm}0.073$&
   $2.538{\pm}0.065$&$2.342{\pm}0.060$&  $-5.56{\pm}0.21$\\  
M 13   & -1.39&-1.18& $ 3.189{\pm}0.086$&$2.987{\pm}0.079$&
   $2.661{\pm}0.069$&$2.421{\pm}0.063$&  $-5.03{\pm}0.21$\\  
M 3    & -1.34&-1.16& $ 3.355{\pm}0.092$&$3.126{\pm}0.086$&
   $2.768{\pm}0.071$&$2.514{\pm}0.063$&  $-4.68{\pm}0.21$\\    
M 5    & -1.11&-0.90& $ 3.310{\pm}0.092$&$3.079{\pm}0.085$&
   $2.694{\pm}0.076$&$2.380{\pm}0.070$&  $-4.81{\pm}0.20$\\  
\\
${\omega}$~Cen& -1.60 & -1.39 &3.202${\pm}0.030$&2.988${\pm}0.030$&
2.648${\pm}0.030$&2.402${\pm}0.030$&-5.03${\pm}0.20$ \\
\\
\hline
\\
\end{tabular}
\end{center}
\end{table*}

\begin{table*}
\begin{center}
\caption[]{RGB {\rm (J\--H)$_0$} colours at fixed magnitudes {\rm (M$_H$=-5.5,
-5, -4, -3)} and H absolute magnitude at constant {\rm (J\--H)$_0$}
for the observed GCs.} 
\label{jh}
\begin{tabular}{lccccccc}
\hline\hline
\\
Name & $[Fe/H]_{CG97}$ & $[M/H]$ & $(J-H)_0^{-5.5}$ & $(J-H)_0^{-5}$ &
$(J-H)_0^{-4}$ &$(J-H)_0^{-3}$& $M_H^{(J\--H)_0=0.7}$\\
\\
\hline
\\
M 15   & -2.12&-1.91 & $0.579{\pm}0.02$ & $0.554{\pm}0.02$ & $0.512{\pm}0.02$ &
 $0.476{\pm}0.01$& $-7.81{\pm}0.29$ \\
M 30   & -1.91&-1.71 & $0.563{\pm}0.01$ & $0.533{\pm}0.01$ & $0.484{\pm}0.01$ &
 $0.447{\pm}0.01$& $-7.63{\pm}0.18$ \\
NGC 6752  & -1.42&-1.21 & $0.658{\pm}0.01$ & $0.630{\pm}0.01$ & $0.579{\pm}0.01$ &
 $0.534{\pm}0.01$& $-6.28{\pm}0.20$ \\
NGC 362   & -1.15&-0.99 & $0.734{\pm}0.01$ & $0.700{\pm}0.01$ & $0.639{\pm}0.01$ &
 $0.585{\pm}0.01$& $-5.00{\pm}0.19$  \\
NGC 288   & -1.07&-0.85 & $0.759{\pm}0.01$ & $0.729{\pm}0.01$ & $0.660{\pm}0.01$ &
 $0.598{\pm}0.01$& $-4.60{\pm}0.18$ \\
NGC 6380  & -0.87&-0.68 & 0.762${\pm}0.04$ & 0.722${\pm}0.04$&0.649${\pm}0.04$ &
0.585${\pm}0.03$& $-4.70{\pm}0.26$ \\
47 Tuc & -0.70&-0.59 & $0.821{\pm}0.02$ & $0.777{\pm}0.01$ & $0.697{\pm}0.01$ &
 $0.628{\pm}0.01$& $-4.04{\pm}0.13$ \\
NGC 6342  & -0.71&-0.53 & $0.856{\pm}0.04$ & $0.807{\pm}0.04$ & $0.719{\pm}0.04$ &
 $0.644{\pm}0.04$& $-3.77{\pm}0.25$  \\
NGC 6441  & -0.68&-0.52 & 0.841${\pm}0.04$&0.794${\pm}0.04$ &0.713${\pm}0.04$&
0.646${\pm}0.03$& $-3.84{\pm}0.28$  \\
NGC 6624  & -0.63&-0.48 & $0.839{\pm}0.04$ & $0.797{\pm}0.04$ & $0.730{\pm}0.04$ &
 $0.657{\pm}0.03$& $-3.71{\pm}0.28$  \\
NGC 6440  & -0.49&-0.40 & $0.867{\pm}0.04$ & $0.823{\pm}0.04$ & $0.746{\pm}0.04$ &
 $0.683{\pm}0.03$& $-3.29{\pm}0.32$  \\
\\
\hline
\\
\end{tabular}
\end{center}
\end{table*}

\begin{table*}
\begin{center}
\caption[]{RGB {\rm (V\--H)$_0$} colours at fixed magnitudes {\rm (M$_H$=-5.5,
-5, -4, -3)} and H absolute magnitude at constant {\rm (V\--H)$_0$} colour
for the observed GCs.} 
\label{vh}
\begin{tabular}{lccccccc}
\hline\hline
\\
Name & $[Fe/H]_{CG97}$ & $[M/H]$ & $(V-H)_0^{-5.5}$ & $(V-H)_0^{-5}$ &
$(V-H)_0^{-4}$ &$(V-H)_0^{-3}$& $M_H^{(V\--H)_0=3}$\\
\\
\hline
\\
M 15   & -2.12	&    -1.91 & $2.815{\pm}0.11$ & $2.674{\pm}0.11$ &
 $2.430{\pm}0.11$ & $2.224{\pm}0.11$& $-6.10{\pm}0.35$ \\
M 30   & -1.91	&    -1.71 & $3.019{\pm}0.09$ & $2.814{\pm}0.08$ &
 $2.495{\pm}0.07$ & $2.274{\pm}0.06$& $-5.46{\pm}0.19$ \\
NGC 6752  & -1.42	&    -1.21 & $3.169{\pm}0.09$ & $2.952{\pm}0.08$ &
 $2.592{\pm}0.07$ & $2.319{\pm}0.06$& $-5.11{\pm}0.19$ \\
NGC 362   & -1.15	&    -0.99 & $3.246{\pm}0.08$ & $3.065{\pm}0.07$ &
 $2.737{\pm}0.07$ & $2.451{\pm}0.07$& $-4.81{\pm}0.21$ \\
NGC 288   & -1.07	&    -0.85 & $3.242{\pm}0.07$ & $3.080{\pm}0.07$ &
 $2.785{\pm}0.07$ & $2.520{\pm}0.06$& $-4.74{\pm}0.23$ \\
NGC 6380  & -0.87& -0.68 &3.880${\pm}0.28$&3.547${\pm}0.27$&
3.047${\pm}0.27$&2.717${\pm}0.26$& $-3.87{\pm}0.38$ \\
47 Tuc & -0.70	&    -0.59 & $4.012{\pm}0.10$ & $3.630{\pm}0.09$ &
 $3.080{\pm}0.07$ & $2.736{\pm}0.06$& $-3.81{\pm}0.17$ \\
NGC 6342  & -0.71	&    -0.53 & $3.975{\pm}0.28$ & $3.641{\pm}0.27$ &
 $3.082{\pm}0.27$ & $2.659{\pm}0.26$& $-3.82{\pm}0.12$ \\
NGC 6441  & -0.68& -0.52 &4.175${\pm}0.31$&3.720${\pm}0.27$ &
3.043${\pm}0.27$&2.624${\pm}0.26$& $-3.92{\pm}0.30$ \\
NGC 6624  & -0.63	&    -0.48 & $4.177{\pm}0.28$ & $3.810{\pm}0.28$ &
 $3.216{\pm}0.27$ & $2.785{\pm}0.26$& $-3.54{\pm}0.23$ \\
NGC 6440  & -0.49	&    -0.40 & $4.186{\pm}0.31$ & $3.683{\pm}0.29$ &
 $3.001{\pm}0.27$ & $2.647{\pm}0.26$& $-4.00{\pm}0.30$ \\
\\
\hline
\\
\end{tabular}
\end{center}
\end{table*}

%
\section*{Acknowledgments}
Part of the data analysis has been performed with the software developed by P.
Montegriffo at the Osservatorio Astronomico di Bologna (INAF).
This publication makes use of data products from the Two Micron All Sky Survey,
which is a joint project of the University of Massachusetts and Infrared
Processing and Analysis Center/California Institute of Technology, founded by
the National Aeronautics and Space Administration and the National Science
Foundation.
The financial support by the Agenzia Spaziale Italiana (ASI) and the Ministero
dell'Istruzione, Universit\'a e Ricerca (MIUR) is kindly acknowledged.
\section*{Appendix}

\appendix

In this appendix we report all the relations linking the photometric indices
defined in the paper as a function of the cluster metallicity in both CG97 
and global scale.\\

{\rm (J\--K)$_0$} colours at fixed M$_K=(-5.5, -5, -4 , -3)$ magnitudes:\\

\begin{equation}
 (J-K)^{M_K=-5.5}_0=0.22[Fe/H]_{CG97}+1.14   
\end{equation} 

\begin{equation}
  (J-K)^{M_K=-5}_0=0.20[Fe/H]_{CG97}+1.06 
\end{equation} 

\begin{equation}
(J-K)^{M_K=-4}_0=0.16[Fe/H]_{CG97}+0.93
\end{equation} 

\begin{equation}
(J-K)^{M_K=-3}_0=0.13[Fe/H]_{CG97}+0.83
\end{equation} 

\begin{equation}
(J-K)^{M_K=-5.5}_0=0.23[M/H]+1.11
\end{equation} 

\begin{equation}
(J-K)^{M_K=-5}_0=0.21[M/H]+1.04
\end{equation} 

\begin{equation}
(J-K)^{M_K=-4}_0=0.17[M/H]+0.92
\end{equation} 

\begin{equation}
(J-K)^{M_K=-3}_0=0.14[M/H]+0.81
\end{equation}

{\rm (V\--K)$_0$} colours at fixed M$_K=(-5.5, -5, -4 , -3)$ magnitudes:\\

\begin{equation}
(V\--K)^{M_K=-5.5}_0=0.90[Fe/H]^2_{CG97}+3.30[Fe/H]_{CG97}+5.98                           
\end{equation} 
  
\begin{equation}
(V-K)^{M_K=-5}_0=0.34[Fe/H]^2_{CG97}+1.50[Fe/H]_{CG97}+4.49                   
\end{equation} 
 
\begin{equation}
(V-K)^{M_K=-4}_0=0.38[Fe/H]_{CG97}+3.29
\end{equation} 

\begin{equation}
(V-K)^{M_K=-3}_0=0.26[Fe/H]_{CG97}+2.87
\end{equation} 

\begin{equation}
(V-K)^{M_K=-5.5}_0=1.10[M/H]^2+3.52[M/H]+5.77
\end{equation} 

\begin{equation}
(V-K)^{M_K=-5}_0=0.41[M/H]^2+1.60[M/H]+4.37
\end{equation} 

\begin{equation}
(V-K)^{M_K=-4}_0=0.40[M/H]+3.23
\end{equation} 

\begin{equation}
(V-K)^{M_K=-3}_0=0.28[M/H]+2.83
\end{equation}

{\rm (J\--H)$_0$} colours at fixed M$_H=(-5.5, -5, -4 , -3)$ magnitudes:\\

\begin{equation}
(J-H)^{M_H=-5.5}_0=0.20[Fe/H]_{CG97}+0.97
\end{equation}

\begin{equation}
(J-H)^{M_H=-5}_0=0.19[Fe/H]_{CG97}+0.92
\end{equation}

\begin{equation}
(J-H)^{M_H=-4}_0=0.16[Fe/H]_{CG97}+0.82
\end{equation}

\begin{equation}
(J-H)^{M_H=-3}_0=0.14[Fe/H]_{CG97}+0.74
\end{equation}

\begin{equation}
(J-H)^{M_H=-5.5}_0=0.21[M/H]+0.94
\end{equation}

\begin{equation}
(J-H)^{M_H=-5}_0=0.20[M/H]+0.90
\end{equation}

\begin{equation}
(J-H)^{M_H=-4}_0=0.17[M/H]+0.80
\end{equation}

\begin{equation}
(J-H)^{M_H=-3}_0=0.15[M/H]+0.72
\end{equation}

{\rm (V\--H)$_0$} colours at fixed M$_H=(-5.5, -5, -4 , -3)$ magnitudes:\\

\begin{equation}
(V-H)^{M_H=-5.5}_0=0.76[Fe/H]^2_{CG97}+2.81[Fe/H]_{CG97}+5.50
\end{equation}

\begin{equation}
(V-H)^{M_H=-5}_0=0.53[Fe/H]^2_{CG97}+2.08[Fe/H]_{CG97}+4.77
\end{equation}

\begin{equation}
(V-H)^{M_H=-4}_0=0.44[Fe/H]_{CG97}+3.30
\end{equation}

\begin{equation}
(V-H)^{M_H=-3}_0=0.36[Fe/H]_{CG97}+2.92
\end{equation}

\begin{equation}
(V-H)^{M_H=-5.5}_0=0.89[M/H]^2+2.89[M/H]+5.23
\end{equation}

\begin{equation}
(V-H)^{M_H=-5}_0=0.66[M/H]^2+2.22[M/H]+4.61
\end{equation}

\begin{equation}
(V\--H)^{M_H=-4}_0=0.46[M/H]+3.24
\end{equation}

\begin{equation}
(V-H)^{M_H=-3}_0=0.37[M/H]+2.87
\end{equation}

M$_K$ magnitudes at fixed (J\--K)$_0=0.7$ and (V\--K)$_0=3$ colours:\\

\begin{equation}
 M^{(J-K)_0=0.7}_K=2.09[Fe/H]_{CG97}-1.16
\end{equation}

\begin{equation}
M^{(V-K)_0=3}_K=1.37[Fe/H]_{CG97}-2.84
\end{equation}

\begin{equation}
M^{(J-K)_0=0.7}_K=2.22[M/H]-1.38
\end{equation}

\begin{equation}
M^{(V-K)_0=3}_K=1.44[M/H]-3.03
\end{equation}

M$_H$ magnitudes at fixed (J\--H)$_0=0.7$ and (V\--H)$_0=3$ colours:\\

\begin{equation}
M^{(J-H)_0=0.7}_H=2.90[Fe/H]_{CG97}-1.87
\end{equation}

\begin{equation}
M^{(V-H)_0=3}_H=1.47[Fe/H]_{CG97}-2.90
\end{equation}

\begin{equation}
M^{(J-H)_0=0.7}_H=3.05[M/H]-2.23
\end{equation}

\begin{equation}
M^{(V-H)_0=3}_H=1.55[M/H]-3.08
\end{equation}

The RGB slope:\\

\begin{equation}
[Fe/H]_{CG97}=-22.21(slope_{RGB})-2.80
\end{equation}

\begin{equation}
[M/H]=-20.83(slope_{RGB})-2.53
\end{equation}

\label{lastpage}

\end{document}